\documentclass[a4paper]{easychair}
\usepackage[utf8]{inputenc}
\usepackage{booktabs}
\usepackage{lstsemantic}
\usepackage{lstlangmizar}
\usepackage{bussproofs}
\hypersetup{
  pdftitle={Escape to Mizar from ATPs},
  pdfauthor={Jesse Alama},
  pdfkeywords={automated theorem proving, mizar, interactive theorem proving, automated reasoning, resolution, natural deduction},
  colorlinks=true,
  linkcolor=black,
  citecolor=black,
}

\let\phi=\varphi

\newcommand{\systemname}[1]{\textsf{#1}}

\newcommand{\mizar}[0]{\systemname{Mizar}}
\newcommand{\systemontptp}[0]{\systemname{SystemOnTPTP}}
\newcommand{\otttomiz}[0]{\systemname{ott2miz}}
\newcommand{\otter}[0]{\systemname{Otter}}
\newcommand{\isabelle}[0]{\systemname{Isabelle}}

\newcommand{\vampire}[0]{\systemname{Vampire}}
\newcommand{\eprover}[0]{\systemname{E}}
\newcommand{\provernine}[0]{\systemname{Prover9}}
\newcommand{\ivy}[0]{\systemname{Ivy}}

\newcommand{\toolname}[1]{\textsl{#1}}
\newcommand{\epclextract}[0]{\toolname{epclextract}}

\usepackage{xcolor}

\theoremstyle{remark}

\newtheorem{example}{Example}

\newcommand{\coloneqq}[0]{\mathrel{\mathop:}=}

\newcommand{\bottom}[0]{\perp}

\newcommand{\tptpproblemurl}[1]{http://www.cs.miami.edu/~tptp/cgi-bin/SeeTPTP?Category=Problems\&Domain=GRA&File={#1}.p}
\newcommand{\tptpproblemlink}[1]{\href{\tptpproblemurl{#1}}{\texttt{#1}}}
\newcommand{\tptpaxiomurl}[1]{http://www.cs.miami.edu/~tptp/cgi-bin/SeeTPTP?Category=Axioms\&File={#1}.ax}

\usepackage{mathabx}

\title{Escape to \mizar{} from ATPs}
\titlerunning{Escape to \mizar{} from ATPs}

\author{Jesse Alama\thanks{Supported by the ESF research project
    \emph{Dialogical Foundations of Semantics} within the ESF
    Eurocores program \emph{LogICCC} (funded by the Portuguese Science
    Foundation, FCT LogICCC/0001/2007).  Research for this paper was
    partially done while a visiting fellow at the Isaac Newton
    Institute for the Mathematical Sciences in the programme
    `Semantics \& Syntax'.  Josef Urban inspired this project and
    provided many helpful suggestions.  Artur Korni{\l}owicz clarified
    some important details of \mizar{} proofs.}\\\affiliation{Center
    for Artificial Intelligence}\\\affiliation{New University of
    Lisbon}\\\affiliation{Portugal}\\\affiliation{\url{j.alama@fct.unl.pt},
    \url{http://centria.di.fct.unl.pt/~alama/}}} \authorrunning{Alama}

\begin{document}

\maketitle

\begin{abstract}
  We announce a tool for mapping \eprover{} derivations to \mizar{}
  proofs.  Our mapping complements earlier work that generates
  problems for automated theorem provers from \mizar{} inference
  checking problems.  We describe the tool, explain the mapping, and
  show how we solved some of the difficulties that arise in mapping
  proofs between different logical formalisms, even when they are
  based on the same notion of logical consequence, as \mizar{} and
  \eprover{} are (namely, first-order classical logic with identity).
\end{abstract}

\section{Introduction}\label{sec:intro}

The problem of
generating a mapping
between proofs in
different formats is
an important
research problem.
Proofs coming from a
many sources can be
found today.  There
are about as many
implemented proof
formats as there are
different systems
for interactive and
automated theorem
proving, not to
mention the ``pure''
proof formats coming
from mathematical
logic.  Even within
the latter we find a
plethora of
possibilities.  If
we pick a
Hilbert-style
system, there is a
choice about which
axioms and rules of
inference to pick.
Even natural
deduction comes in a
number of shapes:
Jáskowski, Gentzen,
Fitch,
Suppes\dots~\cite{pelletier1999brief}.
It seems likely that
as the use of proof
systems grows we
will need to have
better tools for
mapping between
different; this need
has been recognized
for
decades~\cite{wos1990problem,andrews1991more},
and it still seems
we have some way to
go.  % Indeed, prior to any automated transformation of
% one proof into another, we need to know, given a stream of characters
% which is the output of a theorem prover, where the proof even begins
% and where it ends; the well-known SZS dataform standards~\cite{sutcliffe2008szs} helps to
% impose some order.  It seems we have some way to go even here.  (How
% can we transform proofs if we don't even know how to extract them from
% the output of a theorem prover session?)

This paper discusses the problem of transforming derivations output by
the \eprover{} automated theorem prover into \mizar{} texts.

\mizar{} is a language for writing mathematical texts in a ``natural''
style.  It features a kind of natural deduction proof language.  The
library of knowledge formalized in \mizar{}, the \mizar{} Mathematical
Library (MML), is quite advanced, going from the axioms of set theory
to graduate-level pure mathematics.  For the purposes of this paper we
are not interested in the MML.  Instead, we view \mizar{} as a
language and a suite of tools for carrying out arbitrary reasoning in
first-order classical logic.

% The MML begins with set theory axioms,
% with (essentially) only equality and set membership $\in$ as
% primitives, and proceeds to define everything.  Since \mizar{}
% enforces the extension principle (every new function needs to have its
% associated existence and uniqueness condition proved), the whole
% library can be seen as a highly articulated extension by definitions
% of Tarski-Grothendieck set theory.

% We are interested in mapping more or less arbitrary \eprover{}
% derivations into corresponding \mizar{} articles.

% The articles is the main unit of \mizar{} text.  It consists of an
% initial environment specifying the background knowledge that will be
% needed for the development, then proceeds to define new concepts and
% prove theorems.  We construct, for each \eprover{} derivation that
% meets (meeting certain criteria, to be discussed later), a \mizar{}
% text that expresses the same proof expressed by the \eprover{}
% derivation.

Our work is available at
\begin{quotation}
  \url{https://github.com/jessealama/tptp4mizar}
\end{quotation}

Related work is
discussed in
Section~\ref{sec:related-work}.
Section~\ref{sec:translating-tptp}
discusses an
important
preliminary exercise
to mapping
derivations, and
which is perhaps
already of interest:
mapping an arbitrary
TPTP problem (not
necessarily
derivations) into a
corresponding
\mizar{} article.
The generated
\mizar{} text has
the same flat
structure as initial
TPTP problem from
which it comes.
Section~\ref{sec:translating-derivations}
is the heart of the
paper; it discusses
in detail
translation from
\eprover{}
derivations to
\mizar{} proofs.
Because of the
fine-grained level
of detail offered by
\eprover{} and the
simple multi-premise
``obvious
inference'' rule of
\mizar{}, the
mapping is more or
less
straightforward,
save for
\emph{skolemization}
and
\emph{resolution},
neither of which
have direct
analogues in ``human
friendly'' \mizar{}
texts.
Skolemization is
discussed in
Section~\ref{sec:skolemization}
and our treatment of
resolution is
discussed
in~\ref{sec:resolution}.
The problem of
making the generated
\mizar{} texts more
humanly
comprehensible is
discussed in
Section~\ref{sec:compressing}.
Section~\ref{sec:conclusion}
concludes and
proposes
applications and
further
opportunities for
development.
Appendix~A is a
complete example of
a text (a solution
to the Dreadbury
Mansion puzzle found
by \eprover{},
translated to
\mizar) produced by
our translation.

\section{Related work}\label{sec:related-work}

In recent years
there is an interest
in adding automation
to interactive
theorem proving
systems.  An
important challenge
is to make sense, at
the level of the
interactive theorem
prover, of the
solution produced by
external automated
reasoning tools.
Such \emph{proof
  reconstruction}
has been done for
\isabelle/HOL~\cite{paulson2007reconstruction}.
There, the problem
of finding an
\isabelle{}/HOL text
suitable for solving
an inference problem
$P$ is done as
follows:
\begin{enumerate}
\item Translate $P$
  to a first-order
  theorem proving
  problem $P^{*}$.
\item Solve $P^{*}$
  using an automated
  theorem prover,
  yielding solution
  $S^{*}$.
\item Translate
  $S^{*}$ into a
  \isabelle/HOL
  text, yielding a
  solution $S$ of
  the original
  problem.
\end{enumerate}
The work described in this paper could be used to provide a similar
service for \mizar.  It is interesting to note that in the case of
\mizar{} the semantics of the source logic and the logic of the
external theorem prover are the same: first-order classical logic with
identity.  In the \isabelle/HOL case, at step (1) there is a potential
loss of information because of a mismatch of \isabelle/HOL's logic and
the logic of the ATPs used to solve problems (which may not in any
case matter at step~(3)).  In the \mizar{} context, two-thirds (steps
(1) and (2)) of the problem has been solved~\cite{rudnicki2011escape};
our work was motivated by that paper.  Steps toward (3) have been taken
in the form of Urban's \otttomiz{}\footnote{See its homepage
  \url{https://github.com/JUrban/ott2miz} and its announcement
  \url{http://mizar.uwb.edu.pl/forum/archive/0306/msg00000.html} on
  the \mizar{} users mailing list.}.  In fact, more than 2/3 of the
problem is solved.  Our work here builds on \otttomiz{} by accounting
for the clause normal form transformation, rather than starting with
the clause normal form of a problem. Our translated proofs thus start
with (the \mizar{} form of) the relevant initial formulas, which
arguably improves the readability of the proofs.  Moreover, our tool
works with arbitrary TPTP problems and TSTP derivations (produced by
\eprover), rather than with \otter{} proof objects.  The restriction
to \eprover{} is not essential; there is no inherent obstacle to
extending our work to handle TSTP derivations produced by other
automated theorem provers, provided that these derivations are
sufficiently detailed, like \eprover{}'s.  One must acknowledge, of
course, that providing high-quality, fine-grained proof objects is a
challenging practical problem for automated theorem provers.

% The work described
% here could be used to map $L$ into a \mizar{} text, thereby providing
% a service for \mizar{} similar to the one now available to
% \isabelle{}/HOL users.

To account for the clausal normal form transformation, one needs to
deal with skolemization.  This is a well-known issue in discussions
surrounding proof objects for automated theorem
provers~\cite{denivelle2002extraction}.  Interestingly, our method for
handling skolemization is quite analogous to the handling of
quantifiers in the problem opposite ours, namely, converting \mizar{}
proofs to TSTP derivations~\cite{urban2008atp} in the setting of MPTP
(\mizar{} Problems for Theorem Provers)~\cite{urban2006mptp}.  There,
Henkin-type implications are a natural solution to the problem of
justifying a substitution instance of a formula given that its
generalization is justified.  Our justification of skolemization steps
is virtually the same as this; see Section~\ref{sec:skolemization} for
details.

An export and
cross-verification
of \mizar{} proofs
by ATPs has been
carried
out~\cite{urban2008atp}.
Such work is an
inverse of ours
because it goes from
\mizar{} proofs to
ATP problems.

We do not intend to
enter into a
discussion about the
proof identity
problem.  For a
discussion, see
Do{\v{s}}en~\cite{dosen-proof-identity}.
Certainly the
intension behind the
mapping is to
preserve whatever
abstract proof
expressed by the
\eprover{}
derivation.  That
the \eprover{}
derivation and the
\mizar{} text
generated from it
are isomorphic will
be clarified by the
discussion below of
the translation
algorithm.  Mapping
such as the one
discussed here can
help contribute to a
concrete
investigation of the
proof identity
problem, which in
fact motivates the
project reported
here.  The reader
need not share the
author's interest in
the proof identity
problem to
understand what
follows.

It is well-known
that derivations
carried out in
clause-based calculi
(such as resolution
and kindred methods)
tend to be difficult
to understand, if
not downright
inscrutable.  An
important problem
for the automated
reasoning community
for many years is to
find methods whereby
we can understand
machine-discovered
proofs, such as
resolution
refutations.  One
approach to this
problem is to map
resolution
derivations into
natural deduction
proofs.  Much work
has been done in
this
direction~\cite{miller1984expansion,miller1987compact,felty1987proof,egly1997structuring,lingenfelder1989structuring,meier2000tramp}.
The transformations
we employ are rather
simple.  Because of
the coarseness of
\mizar{}'s proof
apparatus (there is
essentially only one
rule of inference
that subsumes most
of the traditional
introduction and
elimination rules of
natural deduction),
we need not be
concerned with a
translation that
preserves the fine
structure of an
\eprover{}
derivation.  To
``clean up'' the
generated text, we
take advantage of
the various proof
``enhancers''
bundled with the
standard \mizar{}
distribution~\cite[\S
4.6]{grabowski2010mizar}.
These enhancers
suggest compressions
of a \mizar{} text
that make it more
parsimonious while
preserving its
semantics.  In the
end, though, it
would seem that the
judgment of whether
an ``enhanced''
\mizar{} text is the
best representative
of a resolution
proof is something
that has to be left
to the reader.

\section{Translating TPTP problems into \mizar{} texts}\label{sec:translating-tptp}

In this section we describe a method for generating a \mizar{} text
from an arbitrary (first-order) TPTP problem~\cite{sutcliffe2009tptp}.
TPTP problems are not themselves derivations, so this mapping is not
the heart of our work.  However, it was an important first step to
mapping derivations to \mizar{} proofs because it revealed some
difficulties that had to be solved in the translation of formulas part
of the mapping of derivations to \mizar{} proofs.  The next section is
devoted to the proof mapping problem.

TPTP is a language for specifying automated reasoning problems.  One
states some axioms and definitions, and perhaps a conjecture.
Although TPTP has in recent years been extended to support various
extensions of the language of first-order logic, we are interested in
this paper only in the first-order part of TPTP.

To construct a \mizar{} text from a TPTP problem, one first identifies
the function and predicate symbols of the TPTP problem and creates a
\emph{environment} for the text.  This step is necessary because
\mizar{} is a richer language than TPTP.  Given a well-formed TPTP
file, one can simply determine, for each symbol appearing in it,
whether it is a function or a predicate, and what it's arity is.
Since (at the time of writing) TPTP focuses only on the case of
one-sorted first-order logic, there is no issue about the sorts of the
arguments and values.  The language of \mizar{}, on the other hand,
permits overloading of various kinds and has (dependent) types.  There
is no issue of inferring from a purported \mizar{} text what the
predicate and function symbols are.  To implement this complexity,
when working with \mizar{} on specifies in advance its so-called
environment.  The environment provides the necessary information to
make sense of the text.

Constructing an environment for a \mizar{} text amounts to creating a
handful of XML files.  Normally, one does not develop \mizar{} texts
from scratch but rather builds on some preexisting formalizations.
Since we not interested in using the \mizar{} library, we cannot use
the usual toolchain.  Instead, we create a fresh environment with
respect to which the generated \mizar{} text is sensible.  This
environment gives a meaning to the TPTP problem even if the TPTP
``problem'' is actually a derivation.  Constructing \mizar{} proofs
from \eprover{} derivations (expressed in the TSTP notation) is the
subject of the next section.

\section{Translating \eprover{} derivations into \mizar{} texts}\label{sec:translating-derivations}

This section discusses the main part of our contribution: mapping
\eprover{} derivations to \mizar{} texts.

The input to our procedure is an \eprover{} derivation in TSTP
format~\cite{sutcliffe2006using} (the standard \eprover{} distribution
comes with a tool, \epclextract, which can translate derivations
expressed in \eprover's custom proof language into proofs in the desired format).

The \mizar{} proof is isomorphic to the \eprover{} derivation in the
sense that the premises $P_{\text{\eprover}}$ of the \eprover{}
derivation map to a set $P_{\text{\mizar}}$ of the same cardinality
and the same logical form, and the conclusion $c_{\eprover}$ of the
$\eprover$ derivation maps directly to the sole theorem $c_{\mizar}$
of the \mizar{} text.  The logical content of the two proofs are the
same because \eprover{} and \mizar{} are both based on first-order
classical logic.  Because \eprover{}'s calculus is based essentially
on clauses while \mizar{} works with formulas, some hurdles need to be
overcome when mapping (i)~the part of an \eprover{} derivation dealing
with converting the input problem to clause normal form, and
(ii)~applications of the rule of resolution.  We describe the mapping
and our solution to these difficulties.

As one might expect, the mapping between an \eprover{} derivation,
which operates essentially on clauses, is not a simple one-to-one
mapping of formulas (more precisely, clauses) to formulas.
\eprover{}'s calculus can to a large extent be recognized by \mizar{}
in the sense that most steps in an \eprover{} derivation do map
directly to (single) steps in the generated \mizar{} text.  Two
classes of inferences, though, raises some problems: skolemization and
resolution, which are the heart of a resolution calculus such as the
one behind \eprover{}.

It seems to be a hard AI problem to transform arbitrary resolution
proofs into human-comprehensible natural deductions.  There often
seems to be a artificial ``flavor'' of such proofs that no spice can
overcome.  Still, some simple organizational principles can help to
make the proof more manageable.  (Later in
Section~\ref{sec:compressing} we will see some stronger syntactic and
semantic methods, going beyond the simple structural guidelines we are
about to discuss, for ``enhancing'' the generated proofs even
further.) Section~\ref{sec:overall} discusses the overall organization
of the generated proof.  In Section~\ref{sec:skolemization} we discuss
the skolemization problem.  In Section~\ref{sec:resolution} we discuss
the problem of resolution.

\subsection{Global and local organization of the proof}
\label{sec:overall}

The first batch of transformation do not compress the derivations in
any way: every step in the TSTP derivation appears in the \mizar{}
output.  However, the refutation is ``groomed'' in the following ways:

\begin{enumerate}
\item Linearly order the formulas.

  Unlike TPTP/TSTP problems, where order of formulas is immaterial,
  the order of formulas in \mizar{} has to be coherent.  We
  topologically sort the input ordered in the obvious way (if
  conclusion $A$ uses formula $B$ as a premise, then $B$ should appear
  earlier than $A$) and work with a linear order.

\item Because one can ``reserve'' variables globally in \mizar, one
  can strip away the initial universal prefix of clauses-as-formulas.

  This transformation not only makes the formulas appearing in the
  proof shorter and hence more readable, it helps to keep \mizar{}'s
  \verb+by+ rule of inference aligned with the various clause-oriented
  rules of inference in \eprover{}'s calculus (clauses don't have
  quantifiers).

\item Separate reasoning done among the axioms (establishing lemmas)
  from the application of lemmas toward the derivation of $\bottom$.

  In other words, we distinguish conclusions that depend on the
  conjecture from conclusions that are independent of it.

\item Separate those lemmas that are used in the refutation proper
  from those that not used.  (I.e., distinguish lemmas that are used
  in the refutation proper from the lemmas that are used only to prove
  other lemmas.)
% \item Squeeze formulas that get rendered identically.

%   \eprover's proof output is very detailed.  It contains fine steps
%   such as detailed conversion from input formulas to clause normal
%   form.  During this transformation, one and the same formula may
%   appear multiple times depending on whether it is being considered as
%   a clause or as a formula.  Input literals (atomic formulas or
%   negations of atomic formulas) can even appear three times in the
%   output as it is being considered in different ways.  Because
%   \mizar{} texts work with formulas, these differences are immaterial
%   for our purposes.  We could have left these duplicate occurences in
%   the text, but we elected to trim them.  We implemented a simple
%   preprocessing step that squeezes multiple occurrences of formulas
%   into unique occurrences.
\end{enumerate}

Step~(1) is strongly necessary because if a conclusion is drawn in a
\mizar{} text from a premise that has not yet been introduced, this is
a fatal error.  Step~(2) is needed for a deeper reason: if we were to
deal always with explicit universal closures of formulas, we would
quickly start to outstrip the notion of obvious inference on which
\mizar{} is based.  Steps~(3)--(5) are not necessary; there is nothing
wrong with disregarding those organizational principles.  However,
there is a cost: abandoning them results in an undifferentiated,
disorganized melange of inferences, a mere ``print out'' in \mizar{}
form of the \eprover{} derivation.

% \mizar{}'s proof checker is obviously much weaker than \eprover{}.
% Moreover, even though \mizar{} can verify most of \eprover's
% fine-grained proof steps, some steps are not verified by \mizar{} as
% they stand.  \mizar{} needs some help:

A refutation starts with some axioms, a conjecture, and proceeds by
negating the conjecture formula and deriving $\bottom$ by reasoning
with the axioms and the negation of the conjecture.  \mizar{} texts in
the \mizar{} Mathematical Library, on the other hand, if read at their
toplevel, are intended to be consistent: given some axioms and lemmas,
one states theorems.  The \emph{proofs} of these lemmas and theorems
may use proof by contradiction, but that is done inside a proof block,
outside of which any contradictory assumptions and conclusions derived
therein are no longer ``accessible''.  However, a TSTP representation
of a refutation is a flat sequence of formulas ending with a
contradiction: the axioms, the conjecture, the negation of the
conjecture, and conclusions drawn among the axioms and the negation of
the conjecture all at the same level.

To capture the spirit of proof by contradiction while ensuring that
the toplevel content of the generated \mizar{} article is coherent (or
at least not manifestly incoherent), we refactor \eprover{}
refutations into so-called diffuse reasoning blocks.  We write:
\begin{lstlisting}[language=Mizar]
theorem @$\phi$@
proof
  now
    assume @$\neg \phi$@;
    S1: @$\langle \text{conclusion 1} \rangle$@ by @\dots@;
    S2: @$\langle \text{conclusion 2} \rangle$@ by @\dots@;
    @\dots@
    S@$n$@: @$\langle \text{conclusion $n$} \rangle$@ by @\dots@;
    thus contradiction by @$\mathtt{S}_{a_{1}}$@, @$\mathtt{S}_{a_{2}}$@, @\dots@, @$\mathtt{S}_{a_{m}}$@
  end;
  hence thesis;
end;
\end{lstlisting}
This concludes the discussion of the organization of the generated
\mizar{} proof.

\subsection{Skolemization}\label{sec:skolemization}

\eprover{}'s finely detailed proof output contains not simply the
derivation of $\bottom$ starting from the clause form of the input
formulas.  \eprover{} can also record the transformation of the input
formulas into clause form.  It is important to preserve these
inferences because they give information about what was actually given
to \eprover; throwing away this information strikes us as unwelcome
because one would have to work harder to make sense of the overall
proof.

If we insist on preserving skolemization steps in the \mizar{} output,
then we have a difficulty in accounting for them.  Carrying out this
task is a well-known issue in generating proof
objects~\cite{denivelle2002extraction,denivelle2005translation}.  The
difficulty is that skolem functions are curious creatures in an
interactive setting like \mizar{}'s.  Introducing a function in into a
\mizar{} text requires that the use can prove existence and uniqueness
of its definiens.  But what is the definiens of a skolem function?

We solve the problem by introducing, as part of the environment of an
article (and not in the generated text), a ``definition'' for skolem
functions in the following manner.  To take a simple example, suppose
we have proved $\forall x \exists y \phi$ and we have that $\forall x
\phi [y \coloneqq f(x)]$ is ``derived'' from this, in the sense that
it is is the conclusion of a skolemization step.  We covertly
introduce at this point a new definition:
\[
(\forall x \exists y \phi) \rightarrow \forall x \phi [y \coloneqq f(x)]
\]
This formula does not have the usual shape of an explicit definition
of a function.  One wonders how one would prove existence and
uniqueness for this definiens.  We do not address these problems; in
effect, the above implication is treated as a new axiom.

Our approach seems
defensible to us.
After all,
\eprover{} does not
give a proof that
introducing the
skolem function is
acceptable, so there
is no step in the
\eprover{}
derivation that
would contain the
needed information.
Giving a proof in
\mizar{} that would
justify
skolemization steps
is in fact possible.
One introduces a new
type $\tau{f}$
inhabited by
definition by those
objects that satisfy
the sentence
$\forall x \exists y
\phi$, prove that
the type is
inhabited by
exploiting the fact
that the domain of
interpretation of
any first-order
structure is
non-empty, and
finally defining $f$
outright using
\mizar{}'s built-in
Hilbert choice
operator.  Initial
experiments with
this approach to
skolemization lead
us to turn off this
feature by default
because it
introduces ``noise''
into the \mizar{}
proof.  We know that
skolemization is a
valid
transformation, so
it seems excessive
to us to put an
explicit
justification of
every skolemization
step.

There is one limitation with the current approach to skolemization at
the moment.  We require that all skolemization steps introduce exactly
one skolem function.

\subsection{Resolution}\label{sec:resolution}

Targeting \mizar{} is sensible because of the presence of a single
rule of inference, called \verb+by+, which takes a variable number of
premises.  The intended meaning of an application

\begin{prooftree}
  \AxiomC{$\phi_{1}$, \dots, $\phi_{n}$}
  \RightLabel{$\mathtt{by}$}
  \UnaryInfC{$\phi$}
\end{prooftree}

of \verb+by+ is that $\phi$ is an ``obvious'' inference from premises
$\phi_{1}$, \dots, $\phi_{n}$.  See Davis~\cite{davis1981obvious} and
Rudnicki~\cite{rudnicki1987obvious} for more information about the the
tradition of ``obvious inference'' in which \mizar{} works.  The
implementation in \mizar{} diverges somewhat from these proposals, but
roughly speaking a conclusion is obtained by an ``obvious inference''
from some premises if there is a Herbrand proof of the conclusion in
which we have chosen at most one substitution instance of each
premise.

One important difficulty for mapping arbitrary resolution proofs to
\mizar{} texts is that \mizar{}'s notion of ``obvious inference''
overlaps with various forms of resolution, but is neither weaker nor
stronger than resolution.  The consequence of this is that it is
generally not the case that an application of resolution can be mapped
to a single acceptable application of \mizar{}'s \verb+by+ rule.
Consider the following example:

% A basic form of resolution looks like:
% \begin{prooftree}
%   \AxiomC{$A \vee C$}
%   \AxiomC{$B \vee \neg C$}
%   \RightLabel{Resol}
%   \BinaryInfC{$A \vee B$}
% \end{prooftree}
% Of course, resolution is much more than this;
% see~\cite{leitsch2005resolution} for a thorough presentation.  For the
% purposes of our discussion it is acceptable to identify ``resolution''
% with this simple rule of inference.  The main point is that resolution
% and the \mizar{} notion of obvious inference diverge.

\begin{example}[Non-obvious resolution inference]
  Consider the inference
  \begin{prooftree}
    \AxiomC{$\neg l(x) \vee d(x)$}
    \AxiomC{$\neg l(x) \vee \neg d(x) \vee \neg d(y)$}
    \RightLabel{Resolution}
    \BinaryInfC{$\neg l(x) \vee \neg d(y)$}
  \end{prooftree}
  Here $l$ and $d$ are unary predicate symbols and $x$ and $y$ are
  variables; all formulas should be read as implicitly universally
  quantified.  This application of resolution simply eliminates $d(x)$
  from the premises.

  If we map the two premises and the conclusion of the application of
  resolution to three \mizar{} theorems and attempt justify the mapped
  conclusion simply by appealing by name to the two mapped premises,
  then we are asking to check an application of \verb+by+, as follows:
  \begin{prooftree}
    \AxiomC{$\forall x \left [ \neg l(x) \vee d(x) \right ]$}
    \AxiomC{$\forall x, y \left [ \neg l(x) \vee \neg d(x) \vee \neg d(y) \right ]$}
    \RightLabel{\texttt{by}}
    \BinaryInfC{$\forall x,y \left [ \neg l(x) \vee \neg d(y) \right ]$}
  \end{prooftree}

  The problem here is that we cannot choose a single substitution
  instance of the premises such that we can find a Herbrand derivation,
  and hence the inference is non-obvious even though it is essentially
  (i.e., at the clause level) a single application of propositional
  resolution.

  The reason for the difficulty is that we are making things
  difficulty for ourselves by working at the level of formulas rather
  than clauses.  A solution is available: map the application of
  resolution not to a single application of \mizar{}'s \verb+by+ rule,
  but to a proof:
  \begin{lstlisting}[language=Mizar]
((not l x) or (not d y))
proof
  A: (not l x) or (not d x)              by Premise1;
  B: (not l x) or (not d x) or (not d y) by Premise2;
  thus thesis by A,B;
end;
  \end{lstlisting}
  There is an application of \mizar's \verb+by+ rule at the end, whose
  conclusion is \verb+thesis+, i.e., the formula to be proved at that
  point in the proof.  We solve the problem by reasoning with
  substitution instances of the premises, obtained by taking instances
  of the premises (these are \verb+A+ and \verb+B+, respectively)
  rather than with whole universal formulas.  Note that the
  substitution instances are not built from constants and function symbols, but
  from (fixed) variables.
\end{example}

\subsection{Compressing \mizar{} proofs}\label{sec:compressing}

The ``epicycles'' of resolution notwithstanding, \mizar{} is able to
compress many of \eprover{}'s proof steps: many steps can be combined
into a single acceptable application of \mizar{}'s \verb+by+ rule of
inference.  For example, if $\phi$ is inferred from $\phi^{\prime}$
from variable renaming, and $\phi^{\prime}$ is inferred by an
application of conjunction elimination to $\phi^{\prime\prime}$,
typically in the \mizar{} setting $\phi$ can be inferred from
$\phi^{\prime\prime}$ alone by a single application of \verb+by+.
This is typical for most of the fine-grained rules of \eprover{}'s
calculus: their applications are acceptable according to \mizar{}'s
\verb+by+, and often they can be composed (sometimes multiple times)
while still being acceptable to \verb+by+.  Other rules in \eprover's
proof calculus that can often be eliminated are variable rewritings,
putting formulas into negation normal form, reordering of literals in
clauses (but recall that \mizar{} proofs are written at the level of
full first-order logic, not in a clause language).  More interesting
compressions exploit the gap between ``obvious inference'' and
\eprover{}'s more articulated calculus.

Compressing proofs helps us to get a sense of what the proof is about.
The \mizar{} notion of obvious inference has been tested through daily
work with substantial mathematical proofs for decades, and thus enjoys
a time-tested robustness (though it is not always uncontroversial).
It seems to be an open problem to specify what we mean by the ``true''
or ``best'' view of a proof.  When \mizar{} texts come from \eprover{}
proofs, \mizar{} finds that the steps are usually excessively detailed
(i.e., most steps are obvious) and can be compressed.  On the other
hand, often the whole proof cannot be compressed into a single
application of \verb+by+.  We employ the algorithm discussed
in~\cite{rudnicki2011escape}: a simple fixed-point algorithm is used
to maximally compress a \mizar{} text.  Thus, by repeatedly attempting
to compress the proof until we reach the limits of \verb+by+, we
obtain a more parsimonious presentation of the proof.

Proof compression is not without its pitfalls; if one compresses
\mizar{} proofs too much, the \mizar{} text can become as ``inhuman''
as the resolution proof from which it comes.  This is a well-known
phenomenon in the \mizar{} community.  % Some of the suggestions can
% steer the text away from human ``naturalness''.
% Concerning one such
% enhancer,
% \toolname{relinfer},
% the \mizar{} team
% themselves write:
% \begin{quotation}
%   \small It can exceptionally shorten proofs but it may also result in
%   poorer readability of the text. For example, some sentences which
%   are important for the proof technically are marked as irrelevant
%   steps, but their removal may force the user to repeat the same
%   library reference (or their combination) instead of a potentially
%   useful lemma; or the removal may be accidental in some sense, that
%   is steps which are crucial for human understanding of the idea of a
%   proof, but are still unnecessary for machine (e.g., unwinding
%   definitions---definitional expansions). Here the tendencies to
%   reduce the complexity of the proof can be misleading.~\cite[p.~196]{grabowski2010mizar}
% \end{quotation}
% Thus,
Applying the proof compression tools seems to require a human's
\emph{bon sens}.  % Using machines to try to ``clean up'' the work of
% machines requires supervision!  The main problem is that often
% multiple compressions of the generated \mizar{} text are available.
Experience with texts generated by our translation shows that
often considerable compression is possible, but at the cost of
introducing a new artificial ``scent'' into the \mizar{}
text.  % Our compression algorithm is deterministic: in those cases when

\section{Conclusion and future work}\label{sec:conclusion}

One naturally wants to extend the work here to work with output of
other theorem provers, such as \vampire{}.  There is no inherent
difficulty in that, though it appears that the TSTP derivations output
by \vampire{} contain different information compared to \eprover{}
proofs; the generic transformations described in
Section~\ref{sec:overall} would carry over, but the mapping of
skolemization and resolution steps of
Sections~\ref{sec:skolemization} and~\ref{sec:resolution} will likely need to be
customized for \vampire.

% The relative lack of information given by \vampire{} compared to
% \eprover{} does not reflect a substantial difference between them.
% From a \vampire{} TSTP derivation one can compute, by inspecting the
% dependency relationships between the formulas, which conclusions are
% drawn from reasoning only among the axioms.  From a programmer's
% perspective, \eprover's output is preferable to \vampire's because the
% computation of the class of conclusions drawn only among the
% inferences is trivial in one case (simply inspect the status of a TPTP
% formula) but mildly non-trivial (and therefore more costly) in the
% other (a dependency graph needs to be constructed and consulted to be
% sure that a formula cannot be traced back to the conjecture).

The TPTP language
recognizes
definitions, but
whether an automated
theorem prover
treats them
differently from an
axiom is
unspecified. In
\mizar{},
definitions play a
vital role.  After
all, \mizar{} is
designed to be a
language for
developing
mathematical
theories; only
secondarily is it a
language for
representing
solutions to
arbitrary reasoning
problems, as we are
using it in this
paper.  One
could try to detect
definitions either
by scanning the
problem looking for
formulas that have
the form of
definitions, or, if
the original TPTP
problem is
available, one can
extract the formulas
whose TPTP status is
\verb+definition+.
Such definition
detection and
synthesis has no
semantic effect, but
could make the
generated \mizar{}
texts more
manageable and
perhaps even
facilitate new
compressions.

At the moment the tool simply translates \eprover{} derivations to
\mizar{} proofs.  A web-based frontend to the translator could help to
spur increased usage (and testing) of our system.  One can even
imagine our tool as part of the \systemontptp{}
suite~\cite{sutcliffe2009tptp}.

An important incompleteness of the current solution is the treatment
of equality.  Some atomic equational reasoning steps (specifically,
inferences involving non-ground equality literals) in \eprover{}
derivations can be non-\mizar{}-obvious.  One possible solution is to
use \provernine's \ivy{} proof objects.  \ivy{} derivations provide
some information (namely, which instances of which variables in
non-ground literals) that (at present) is missing from \eprover{}'s
proof object output.

For the sake of
clarity in the
mapping of
skolemization steps
in \eprover{}
derivation to
\mizar{} steps, we
restricted attention
to those \eprover{}
derivations in which
each skolemization
step introduces
exactly one new
skolem function.
The restriction does
not reflect a
weakness of
\mizar{}; it is a
merely technical
limitation and we
intend to remove it.

We have thus completed the cycle started in~\cite{rudnicki2011escape} and returned from ATPs to \mizar{}.
We leave it to the reader to decide whether he wishes to escape again.

\bibliographystyle{plain}
\bibliography{paar}

\newpage\appendix

\section{Pelletier's Dreadbury Mansion Puzzle: From \eprover{} to \mizar{}}

\begin{lstlisting}[language=Mizar,basicstyle=\ttfamily\scriptsize,escapechar=]
Ax1: ex X1 st (lives X1 & killed X1,agatha) by AXIOMS:1;

Ax2: lives X1 implies (X1 = agatha or X1 = butler or X1 = charles) by AXIOMS:2;

Ax3: killed X1,X2 implies hates X1,X2 by AXIOMS:3;

Ax4: killed X1,X2 implies (not richer X1,X2) by AXIOMS:4;

Ax5: hates agatha,X1 implies (not hates charles,X1) by AXIOMS:5;

Ax6: (not X1 = butler) implies hates agatha,X1 by AXIOMS:6;

Ax7: (not richer X1,agatha) implies hates butler,X1 by AXIOMS:7;

Ax8: hates agatha,X1 implies hates butler,X1 by AXIOMS:8;

Ax9: ex X2 st (not hates X1,X2) by AXIOMS:9;

Ax10: not agatha = butler by AXIOMS:10;

S1: killed skolem1,agatha by Ax1,SKOLEM:def 1;

S2: agatha = skolem1 or butler = skolem1 or charles = skolem1 by Ax2,Ax1,SKOLEM:def 1;

S3: not hates agatha,(skolem2 butler) by Ax9,SKOLEM:def 2,Ax8;

S4: hates charles,agatha or skolem1 = butler or skolem1 = agatha by Ax3,Ax1,SKOLEM:def 1,S2;

S5: butler = (skolem2 butler) by S3,Ax6;

S6: not hates butler,butler by Ax9,SKOLEM:def 2,S5;

S7: hates butler,butler or skolem1 = agatha by Ax4,Ax7,Ax1,SKOLEM:def 1,Ax5,S4,Ax6,Ax10;

S8: skolem1 = agatha by S7,S6;

theorem
killed agatha,agatha
proof
  now
    assume S9: not killed agatha,agatha;
    thus contradiction by S1,S8,S9;
  end;
  hence thesis;
end;
\end{lstlisting}
Pelletier's Dreadbury Mansion~\cite{Pel86-JAR} goes as follows:
  \begin{quotation}
    {\small
    Someone who lives in Dreadbury Mansion killed Aunt Agatha.
    Agatha, the butler, and Charles live in Dreadbury Mansion, and
    are the only people who live therein. A killer always hates his
    victim, and is never richer than his victim.  Charles hates no
    one that Aunt Agatha hates. Agatha hates everyone except the
    butler. The butler hates everyone not richer than Aunt
    Agatha. The butler hates everyone Aunt Agatha hates. No one
    hates everyone. Agatha is not the butler.}
  \end{quotation}
  The problem is: Who killed Aunt Agatha?  (Answer: she killed
  herself.) The problem belongs to the TPTP Problem Library (it is
  known there as \tptpproblemlink{PUZ001+1}) and can easily by solved
  by many automated theorem provers.  Above is the result of mapping
  \eprover's solution to a standalone \mizar{} text and then
  compressing it as described in Section~\ref{sec:compressing}.  Two
  skolem functions \verb+skolem1+ (arity~0) and \verb+skolem2+
  (arity~2) are introduced.  There are 10 axioms and 8 steps that do
  not depend depend on the negation of the conjecture
  (\verb+killed agatha,agatha+) This problem is solved essentially by
  forward reasoning from the axioms; proof by contradiction is
  unnecessary, but that is the nature of \eprover's solution.
\end{document}